\documentclass[twocolumn]{autart}    
\usepackage{multicol}                                   
\usepackage{enumerate}
\usepackage{stfloats}
\usepackage{amsmath}
\usepackage{amssymb}
\usepackage{amsfonts}
\usepackage{graphicx}          

\begin{document}

\begin{frontmatter}

\title{Finite-time stabilization control of quantum systems\thanksref{footnoteinfo}} 

\thanks[footnoteinfo]{
This work was supported in part by the National Natural Science Foundation of China under Grant 61873251, Grant 61828303, and Grant 61773370, and in part by the Australian Research Council's Discovery Projects Funding Scheme Under Project DP190101566.
}

\author[AD1]{Sen Kuang}\ead{skuang@ustc.edu.cn},    
\author[AD1]{Xiaoke Guan}\ead{zerogxk@mail.ustc.edu.cn},              
\author[AD2]{Daoyi Dong}\ead{daoyidong@gmail.com}                   

\address[AD1]{Department of Automation, University of Science and Technology of China, Hefei 230027, PR China}  
\address[AD2]{School of Engineering and Information Technology, University of New South Wales, Canberra ACT 2600, Australia}  


\begin{abstract}                          
The finite-time control problem of quantum systems is investigated in this paper. We first define finite-time stability and present a finite-time Lyapunov stability criterion for finite-dimensional quantum systems in coherence vector representation. Then, for two-level quantum systems, we design a continuous non-smooth control law with a state-dependent fractional power and prove the uniqueness of solutions of the system dynamics with the controller via the concept of transversality. By combining the finite-time Lyapunov stability criterion with the homogeneity theory, the finite-time convergence of the system to an eigenstate of its internal Hamiltonian is proved. Numerical results on a spin-1/2 system demonstrate the effectiveness of the proposed finite-time stabilization control scheme.
\end{abstract}

\end{frontmatter}

\section{Introduction}\label{sec1}
Quantum control has been a fundamental task in the development of quantum science and technology. Many classical control and optimization methods have been applied to quantum systems, e.g.,
optimal control \cite{dolde2014high,PhysRevA.95.063418}, Lyapunov control \cite{altafini2007feedback,wang2010analysis,ZHAO20121833,kuang2017rapid,kuang2018lyapunov}, sliding mode control \cite{dong2009sliding,dong2012sliding}, $H^\infty$ control \cite{4625217,XIANG20178},
fault-tolerant control and filtering \cite{7556290,GAO2016125}, and learning control \cite{dong2019learning,niu2019universal}.
In quantum control, a relevant objective is achieving finite-time control, that is, a desired target state is exactly reached under the action of control fields within a finite time. Since finite-time control can demonstrate high control accuracy, fast convergence, and strong robustness to various uncertainties \cite{Bhat2000finite}, it is of particular relevance for the development of high-precision information processing, quantum metrology,  quantum navigation, quantum sensing and quantum radar.

There have been several existing approaches that can be used for finite-time control of quantum systems, e.g., $\pi$-pulse methods \cite{Allen1987Optical}, methods based on Lie group decomposition \cite{D2007introduction,schirmer2001quantum}, and optimal control methods \cite{d2001optimal,boscain2006time}. These methods can achieve good performance for some problems but they also have their own weaknesses. For example, $\pi$ pulses generally are extremely sensitive to pulse areas.
The Lie group decomposition approach can be used to constructively obtain simple control pulses such as square-wave pulses or Gaussian wavepackets. However, it may be difficult to decompose a desired unitary operator in many practical applications and to generalize the approach to open systems. Optimal control methods often suffer from complex numerical or analytical computing \cite{d2001optimal}. For instance, in time optimal control methods \cite{boscain2006time}, the switching time of bang-bang control needs to be exactly determined, which is often a difficult task. In this paper, we aim to develop an effective finite-time control method that is expected to easily design and implement.

In our finite-time control method, a non-smooth control law will be designed. Generally, non-smooth control can complete more complex control tasks than smooth control. For example, to break symmetric topology of the state space and achieve desired global stabilization, several switching control approaches based on state space partition were proposed for open quantum systems under continuous measurement in \cite{mirrahimi2007stabilizing}, \cite{ge2012non} and \cite{liu2016lyapunov}. Considering feedback delay, Ge \emph{et al. }\cite{ge2012quantum} designed a time delay switching controller to compensate for the control-computation time.
To deal with the uncertainties in the system Hamiltonian, two sliding mode control schemes based on unitary control and periodic projective measurements were proposed for quantum systems in \cite{dong2009sliding} and \cite{dong2012sliding}. Ref. \cite{kuang2017rapid} proposed two switching Lyapunov control approaches to achieve rapidly convergent control for two-level quantum systems. It should be pointed out that these control laws belong to discontinuous non-smooth control. In practical applications, they may cause chattering and cannot induce finite-time convergence to the target state. Although a continuous non-smooth control law in saturation form was also designed for two-level systems in \cite{ge2012non}, the finite-time convergence was not considered. Ref. \cite{johnson2017exact} investigated the finite-time stabilization problem of multipartite entangled states for discrete-time Markovian dynamics by dissipative quantum circuits and presented several conditions for finite-time stabilization and robust finite-time stabilization.

In classical control, there have been several well-established methods for the design of continuous non-smooth finite-time controllers, which typically include the adding-a-power-integrator technology \cite{huang2005global,fu2017adaptive}, finite-time homogeneity methods \cite{bhat2005geometric}, and continuous nonsingular terminal sliding mode control \cite{yu2005continuous,yang2013continuous}. Among these methods, the finite-time Lyapunov stability theory and finite-time homogeneity theory form theoretical bases for the analysis and synthesis of the finite-time control problem. Here we will present a finite-time Lyapunov stability criterion for finite-dimensional closed quantum systems and design a continuous non-smooth control law for two-level quantum systems to achieve finite-time convergence to the target state. It is worth mentioning that the control law in this paper is designed in a feedback way and should be implemented in an open-loop way since measurement is not involved in the controller design, just like most existing work on Lyapunov control of quantum systems  (also see \cite{altafini2007feedback}-\cite{kuang2018lyapunov}). Compared with existing methods of finite-time control of quantum systems, our method has the following advantages: (i) it may avoid complicated numerical computing in design, (ii) it may avoid potential chattering due to continuity, and (iii) it can be generalized to high-dimensional or open quantum systems.
In view of the fact that the continuous non-smooth control system may not satisfy the Lipschitz continuity condition at some points, we demonstrate the uniqueness of solutions of the system dynamics via the concept of transversality, which has been used, e.g., in \cite{kawski1989stabilization}. Readers can refer to \cite{filippov2013differential} and \cite{agarwal1993uniqueness} for other methods to show the existence and uniqueness of solutions for general non-smooth systems.

The main contributions of this paper are summarized as follows. First, finite-time stability and a finite-time Lyapunov stability criterion are presented for finite-dimensional quantum systems.
Second, we propose a continuous non-smooth control law with a state-dependent fractional power for two-level quantum systems via the Lyapunov method, which enables the rapid finite-time convergence of the system. Third, we prove the uniqueness of solutions of the system dynamics with the designed finite-time controller using the transversality condition. Finally, the finite-time Lyapunov stability criterion and the homogeneity theory are simultaneously used to prove the finite-time stability of the controlled quantum system, i.e., the target state is reached within a finite time.

This paper is organized as follows. Section \ref{sec2} introduces the quantum control system and presents the definition of finite-time stability for quantum systems and a Lyapunov criterion for finite-time stability. In Section \ref{sec3}, we design a continuous non-smooth controller for two-level quantum systems via the Lyapunov method and prove the uniqueness of solutions of the control system. The finite-time convergence of the system to the target state is proved in Section \ref{sec4}. Section \ref{sec5} presents numerical results to demonstrate the effectiveness of the proposed finite-time control scheme. Conclusions are presented in Section \ref{sec6}.

\textbf{Notation.} Let $\mathbb{R}_+$ be the set of non-negative real numbers, $\nabla$ be a vector differential operator, $\langle\,,\,\rangle$ denote an inner product operation, and $[A,B]$ denote the commutator between $A$ and $B$. The two state vectors $|\psi_1\rangle$ and $|\psi_2\rangle$ satisfying $|\psi_1\rangle=e^{i\phi}|\psi_2\rangle$, $\phi\in[0,2\pi)$ are said to be equivalent, and the set of all state vectors equivalent to $|\psi\rangle$ forms the equivalence class of $|\psi\rangle$. In physics, equivalent state vectors have the same observation meaning, and therefore can be regarded as the same state.

\section{Finite-time stability of quantum systems}\label{sec2}

\subsection{Basic concepts of finite-time stability}\label{sec2.1}
For an $n$-dimensional closed quantum system, its state can be represented by a unit column vector $|\psi\rangle$ in the Hilbert space defined on $\mathbb{C}^n$ and its dynamics obey the Schr\"{o}dinger equation
\begin{equation} \label{e1}
|\dot{\psi}(t)\rangle=\frac{-i}{\hbar}H{|\psi(t)\rangle}=\frac{-i}{\hbar}\Big(H_0+\sum_{k=1}^rH_ku_k\Big)|\psi(t)\rangle,
\end{equation}
where $H_0$ and $H_k$ are the internal and control Hamiltonians of the system, respectively, $\hbar$ is the reduced Planck constant (set as $\hbar=1$ in this paper), and $u_k$ is an external control field to be designed.

In coherence vector representation, the quantum state $|\psi\rangle$ can be written into
\begin{equation}\label{e2}
{{|\psi\rangle}{\langle\psi|}={\xi_0}{\sigma_0}+\frac{1}{2}\sum_{\kappa=1}^{n^2-1}{\xi_\kappa}{\sigma_\kappa}=\frac{I_n}{n}+\frac{1}{2}\sum_{\kappa=1}^{n^2-1}{\xi_\kappa}{\sigma_\kappa}},
\end{equation}
where $\{\sigma_\kappa\}_{\kappa=0}^{n^2-1}$ is an orthogonal basis of the $n\times n$ complex Hermitian matrix space, $\sigma_0=\frac{I_n}{\sqrt{n}}$, and $\xi_0=\frac{1}{\sqrt{n}}$. The real vector $[\xi_1,\ldots,\xi_{n^2-1}]^\mathrm{T}\triangleq s=[\langle\psi|\sigma_1|\psi\rangle,\ldots,\langle\psi|\sigma_{n^2-1}|\psi\rangle]^\mathrm{T}\in\mathbb{R}^{n^2-1}$ is called the coherence vector of $|\psi\rangle$ in the basis $\{\sigma_\kappa\}_{\kappa=0}^{n^2-1}$. The set of all coherence vectors forms the Bloch space $\mathcal{B}(\mathbb{R}^{n^2-1})$ \cite{kimura2003bloch}.

For simplicity, we only consider the case of one control field. In this case, quantum system (\ref{e1}) can be written as \cite{wang2010analysis}
\begin{equation}\label{e3}
{\dot{s}(t)=(A_0+u_1A_1)s(t)},
\end{equation}
where
the $(m,n)$th elements of $A_0$ and $A_1$ are
\begin{gather}
A_0(m,n)=\mathrm{tr}(i H_0\left[\sigma_m,\sigma_n\right]),\label{e5}\\
A_1(m,n)=\mathrm{tr}(i H_1\left[\sigma_m,\sigma_n\right]).\label{e6}
\end{gather}
Assume that the control law $u_1$ in \eqref{e3} is a continuous function of the state $s$. Thus, system (\ref{e3}) can be written as
\begin{equation}\label{e7}
{\dot{s}(t)=f(s(t)),\; s(t)\in\mathcal{B}(\mathbb{R}^{n^2-1})}
\end{equation}
where $f:\mathcal{B}(\mathbb{R}^{n^2-1})\rightarrow\mathcal{B}(\mathbb{R}^{n^2-1})$ is a continuous function defined on $\mathcal{B}(\mathbb{R}^{n^2-1})$.

To illustrate the concept of finite-time stability, we assume that initial time $t=0$ and quantum system (\ref{e7}) has a unique solution in  $\mathcal{B}(\mathbb{R}^{n^2-1})$ for any initial vector $s_0\in\mathcal{B}(\mathbb{R}^{n^2-1})$. We denote this solution as $s(t)$ or $s(t,s_0)$, ($t\geq0$).
Now, we give the definition of finite-time stability for quantum system (\ref{e7}).

\begin{defn}\label{1}
For quantum system (\ref{e7}), the target vector $s_f$ is said to be finite-time stable if for an arbitrarily given initial vector $s_0\in\mathcal{B}(\mathbb{R}^{n^2-1})$, there exists a continuous function $T(s_0):\mathcal{B}(\mathbb{R}^{n^2-1})\rightarrow[0,\infty)$ such that the unique solution $s(t,s_0)$ of system (\ref{e7}) satisfies $\lim_{t\rightarrow T(s_0)}s(t,s_0)=s_f$ and $s(t,s_0)=s_f$ for $t\geq T(s_0)$. $T(s_0)$ is called the settling time associated with $s_0$.
\end{defn}

Here we consider an example to illustrate the definition of finite-time stability.
\begin{exmp}\label{3}
Consider the scalar differential equation
\begin{equation}\label{e13}
{\dot{y}\left(t\right)=-k\mathrm{sign}\left(y\left(t\right)\right)|y\left(t\right)|^\alpha},
\end{equation}
where $\mathrm{sign}(0)=0$, $k>0$, and $\alpha\in(0,1)$.

Since the right-hand side of (\ref{e13}) is continuous everywhere and the local Lipschitz condition is always satisfied outside the origin, system (\ref{e13}) has a unique solution for any initial condition $y_0\in\mathbb{R}$. By direct integration, the solution of system (\ref{e13}) can be obtained as
\begin{equation}\label{e14}
\mu\left(t,y_0\right)=
\left\{
   \begin{array}{lll}
   \mathrm{sign}(y_0)&\!\!\left[|y_0|^{1-\alpha}-k(1-\alpha)t\right]^{\frac{1}{1-\alpha}},\\
       &\quad\quad \big(t<\frac{|y_0|^{1-\alpha}}{k(1-\alpha)},\,y_0\neq0\big)\\
   0,&\quad\quad \big(t\geq\frac{|y_0|^{1-\alpha}}{k(1-\alpha)},\,y_0\neq0\big)\\
   0,&\quad\quad \big(t\geq0,\,y_0=0\big).
   \end{array}
\right.
\end{equation}
It is known from (\ref{e14}) that the settling-time function
is $T(y_0)=\frac{1}{k(1-\alpha)}|y_0|^{1-\alpha}$. The Lyapunov function $V(y)=y^2$ can be used to prove that the origin of system (\ref{e13}) is globally finite-time stable. Here, we omit the proof for brevity.
\end{exmp}

\subsection{Lyapunov theorem for finite-time stability}\label{sec2.2}
We first give a comparison lemma \cite{khalil2002nonlinear}.
\begin{lem}\label{4}
Let $V$ be a Lyapunov function defined on $\mathbb{R}_{+}\times\mathcal{B}(\mathbb{R}^{n^2-1})$ and assume that
$\dot{V}_E(t,m)\leq\gamma(t,V(t,m))$ holds,
where $(t,m)\in\mathbb{R}_{+}\times\mathcal{B}(\mathbb{R}^{n^2-1})$, ${E}$ denotes the differential equation $\dot{x}=F(t,x)$, $\dot{V}_E(t,m)$ represents the time derivative of the Lyapunov function $V$ along the trajectories of $E$, and  $\gamma:\mathbb{R}_{+}\times\mathbb{R}\rightarrow\mathbb{R}$ is a continuous function. Further, assume that the initial value problem $\dot{m}=\gamma(t,m)$ with $m(t_0)=m_0$ has a unique solution $m(t,m_0)$ in the interval $[t_0, T)$, where $0\leq t_0<T\leq+\infty$. Let $x(t)$, $t\in[t_0,T)$ be a solution of $E$ with $V(t_0,x(t_0))\leq m_0$. Then, $V(t,x(t))\leq m(t,m_0)$ holds for every $t\in[t_0,T)$.
\end{lem}

Based on Lemma \ref{4}, we have the following finite-time stability theorem for quantum system (\ref{e7}).
\begin{thm}\label{5}
For quantum system (\ref{e7}), suppose that $s_f$ is the target vector and there exists a continuously differentiable function $V:\mathcal{B}(\mathbb{R}^{n^2-1})\rightarrow\mathbb{R}$ such that the following conditions hold:\\
(i) $V$ is positive definite; \\
(ii) For $s_0\in\mathcal{B}(\mathbb{R}^{n^2-1})$, there exist two positive real numbers $c>0$ and $\alpha\in(0,1)$ such that
\begin{equation}\label{e16}
\dot{V}\left(s\left(t,s_0\right)\right)+c\left(V\left(s\left(t,s_0\right)\right)\right)^\alpha\leq0.
\end{equation}
Then, system (\ref{e7}) is finite-time stable, that is, it converges to the target vector $s_f$ within a finite time. The settling time function $T(s_0)$ satisfies
\begin{equation}\label{e17}
{T\left(s_0\right)\leq{\frac{1}{c\left(1-\alpha\right)}V\left(s_0\right)^{1-\alpha}}}.
\end{equation}
\end{thm}
\begin{pf}
Considering (\ref{e13}) in Example \ref{3} and letting $y(t)=V(s(t,s_0))$ and $k=c$,  we have
\begin{equation}\label{e18}
{\dot{V}\left(s(t,s_0)\right)=-c\left(V\left(s(t,s_0)\right)\right)^\alpha}.
\end{equation}
For $t\in\mathbb{R}_+$ and $s_0\in\mathcal{B}\big(\mathbb{R}^{n^2-1}\big)$, applying Lemma \ref{4} to the differential inequality (\ref{e16}) and the scalar differential equation (\ref{e18}) yields
\begin{equation}\label{e19}
V\left(s\left(t,s_0\right)\right)\leq{\mu\left(t,V\left(s_0\right)\right)},
\end{equation}
where $\mu$ can be written as
\begin{equation}\label{e20}
\mu\left(t,V\left(s_0\right)\right)=\left\{
   \begin{array}{lll}
   [V(s_0)&\!\!^{1-\alpha}-c\left(1-\alpha\right)t]^{\frac{1}{1-\alpha}},\\
        &\big(t<\frac{V\left(s_0\right)^{1-\alpha}}{c\left(1-\alpha\right)},\,s_0\ne s_f\big)\\
   0,&\big(t\geq\frac{V\left(s_0\right)^{1-\alpha}}{c\left(1-\alpha\right)},\,s_0\ne s_f\big)\\
   0,&\big(t\geq0,\,s_0=s_f\big).
   \end{array}
\right.
\end{equation}
Equation (\ref{e20}) means that the right-hand side of \eqref{e19} vanishes when $t\geq\frac{1}{c(1-\alpha)}\left(V(s_0)\right)^{1-\alpha}$ and therefore $V\left(s(t,s_0)\right)=0$, that is,
\begin{equation}\label{e21}
s(t,s_0)=s_f.
\end{equation}
Since $s(t,s_0)$ is continuous, $\inf\{t\in\mathbb{R}_+:s(t,s_0)=s_f\}>0$  for $s_0\in\mathcal{B}(\mathbb{R}^{n^2-1})\setminus s_f$ and $\inf\{t\in\mathbb{R}_+:s(t,s_0)=s_f\}<\infty$ for $s_0\in\mathcal{B}(\mathbb{R}^{n^2-1})$. Let $T(s_0)\triangleq\inf\{t\in\mathbb{R}_+:s(t,s_0)=s_f\}$. According to Definition \ref{1}, system (\ref{e7}) is finite-time stable to the target vector $s_f$. From (\ref{e19})-(\ref{e21}), it is clear that (\ref{e17}) holds.  \quad$\blacksquare$
\end{pf}
Theorem \ref{5} is a Lyapunov criterion for the finite-time stability of quantum system (\ref{e7}). The homogeneity theory also can be used to determine finite-time stability. Several results related to homogeneity are listed in the Appendix.
We will use Theorem \ref{5} and the homogeneity theory to prove the finite-time stability of two-level quantum systems in Section \ref{sec4}.

\section{Finite-time controller design for two-level quantum systems }\label{sec3}
A two-level quantum system can act as a basic information unit in quantum information processing.
In this section, for two-level quantum systems with only one control field, we design the control law $u_1$ in (\ref{e1}) via the Lyapunov method to realize finite-time convergence of the system to an eigenstate $|\psi_f\rangle$ of $H_0$.

We assume that the internal and control Hamiltonians in this case are given as
\begin{equation}\label{e22e23}
H_0=\left[\begin{array}{cc}
 1&0\\
 0&-1
 \end{array}\right],\;
H_1=\left[\begin{array}{cc}
 0&-i\\
 i&0
 \end{array}\right].
\end{equation}
Denote the eigenstates of $H_0$ as $|0\rangle=[1,0]^\mathrm{T}$ and $|1\rangle=[0,1]^\mathrm{T}$ and assume that the target state is $|\psi_f\rangle=|1\rangle$.

We use the following Lyapunov function \cite{shuang2007quantum}
\begin{equation}\label{e25}
{V=1-|{\langle\psi_f|\psi\rangle}|^2}.
\end{equation}
Its first-order time derivative can be calculated as
\begin{equation}\label{e26}
{\dot{V}=-2u_1|{\langle\psi|\psi_f\rangle}|\Im\big[e^{i\angle{\langle\psi|\psi_f\rangle}}{\langle\psi_f|H_1|\psi\rangle}\big]},
\end{equation}
where we define $\angle{\langle\psi|\psi_f\rangle}=0$ when $\langle\psi_f|\psi\rangle=0$. To guarantee $\dot{V}\leq0$, we design a continuous non-smooth control law with a fractional power as
\begin{equation}\label{e27}
{u_1=K\textrm{sign}\left(\phi_\alpha\left({|\psi\rangle}\right)\right)|\phi_\alpha\left({|\psi\rangle}\right)|^\alpha}
\end{equation}
with $K>0$, $\phi_\alpha(|\psi\rangle)=\Im\left[e^{i\angle{\langle\psi|\psi_f\rangle}}{\langle\psi_f|H_1|\psi\rangle}\right]$, and $\alpha\in(0,1)$.

We apply the homogeneity criterion for finite-time stability (see Lemma \ref{A6} in the Appendix) to prove that the controller in (\ref{e27}) can achieve the finite-time stabilization of system (\ref{e1}). To this end, we need to calculate the degree of homogeneity of the system. By expressing a complex number in its exponential form, the controlled quantum state can be written as
\begin{equation}\label{e28}
{{|\psi\rangle}=\left[x_1,x_2\right]^\mathrm{T}=r_{1}e^{i\phi_a}{|0\rangle}+r_{2}e^{i\phi_b}{|1\rangle}},
\end{equation}
where $r_1,r_2\in[0,1]$ and $r_{1}^{2}+r_{2}^{2}=1$. The difference $\phi_b-\phi_a\triangleq\phi$ of the global phase factors $e^{i\phi_a}$ and $e^{i\phi_b}$ is called the relative phase of $|\psi\rangle$. We also define the phase of $x_j$ to be 0 when $x_j=0\;(j=1,2)$.

With \eqref{e28}, (\ref{e25})-(\ref{e27}) can be written as
\begin{gather}
{V=1-|{\langle\psi_f|\psi\rangle}|^2=r_1^2},\label{e33}\\
\dot{V}=-2Kr_{2}|r_1\cos\phi|^{\alpha+1},\label{e34}\\
u_1=K\textrm{sign}\left(r_{1}\cos\phi\right)|r_{1}\cos\phi|^\alpha.\label{e35}
\end{gather}
According to \eqref{e34}, $\dot{V}=0$ implies $r_2=0$ or $r_1=0$ or $\cos\phi=0$. When $r_2=0$ and $\cos\phi\ne0$, the system is in an equivalence class of $|0\rangle=[1,0]^\mathrm{T}$. In this case, it follows from (\ref{e35}) that $u_1\neq0$ and therefore the system state is transferring towards the target state $|\psi_f\rangle$. When $r_1=0$, the system is in the equivalence class of $|\psi\rangle=|\psi_f\rangle$. In this case, it follows from (\ref{e33})-(\ref{e35}) that $V=0$, $\dot{V}=0$, and $u_1=0$. Considering the positive definiteness of $V$ and the negative definiteness of $\dot{V}$, we know that the system will be stabilized in the equivalence class of the target state.
For $\cos\phi=0$ and $r_1\ne0$, we denote the quantum state satisfying these two conditions as $|\psi_q\rangle$ and the corresponding moment as $t_q$. Although $u_1(t_q)=0$ holds in this case, the relative phase $\phi$ will continue evolving under the internal Hamiltonian. This means that there exists $t_1>t_q$ such that $\cos\phi(t)\neq0$ $(t_q<t\leq t_1)$. From (\ref{e35}) we know $u_1(t)\neq0$ $(t_q<t\leq t_1)$. Thus, the system state will keep evolving towards the target state $|\psi_f\rangle$ and will not remain at $|\psi_q\rangle$ forever. That is, all moments $t_q$ form a zero-measure set. Hence, the state $|\psi_q\rangle$ and the moment $t_q$ do not change the stability property of the controlled system.

The system (\ref{e1}) with the controller (\ref{e35}) does not satisfy the Lipschitz continuity condition at some points. Here, we
show the existence and uniqueness of solutions to two-level quantum systems considering a sufficient condition \cite{kawski1989stabilization}.
\begin{thm}\label{6}
Under the action of the controller (\ref{e35}), the two-level quantum system (\ref{e1}) with the Hamiltonians as shown in (\ref{e22e23}) has a unique continuously differentiable solution for every initial state.
\end{thm}
\begin{pf}
The Bloch vector of a two-level quantum system can be represented by $s=(\sin\theta\cos\phi,\sin\theta\sin\phi,\cos\theta)$ in the Bloch spherical coordinate frame (see Fig. \ref{fig1}). Accordingly, the system state can be written as
\begin{equation}\label{e36}
{|\psi(\theta,\phi)\rangle}=\big[\begin{array}{c}
\cos\frac{\theta}{2},\,e^{i\phi}\sin\frac{\theta}{2}\end{array}\big]^\mathrm{T},
\end{equation}
where $\theta\in[0,\pi]$ and $\phi\in[0,2\pi)$ \cite{fano2017twenty}. The relative phase $\phi$ of $|\psi\rangle$ is the angle between the projection of $s$ on the $x-y$ plane and the positive $x-$axis, and $\theta$ is the angle between $s$ and the positive $z-$axis.
\begin{figure}
\begin{center}
\includegraphics[height=4cm]{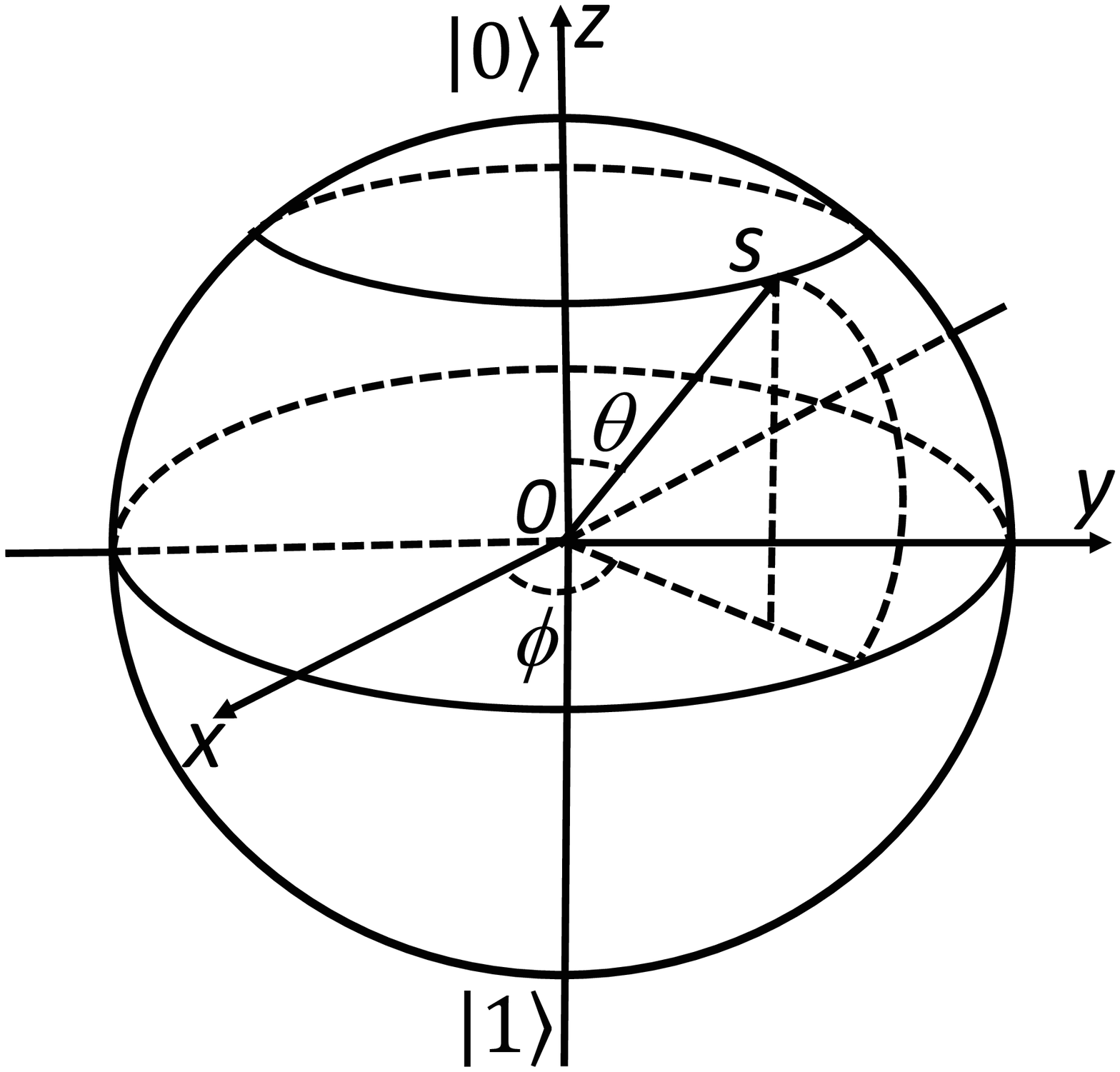}    
\caption{The Bloch vector of a two-level quantum system.}  
\label{fig1}                                 
\end{center}                                 
\end{figure}

For an initial state outside the set $\mathcal{O}=\{|\psi\rangle:\cos\phi=0,\, r_1\ne0\}$, the vector field of system (\ref{e1}) with the controller (\ref{e35}) is Lipschitz everywhere and therefore system (\ref{e1}) has a unique solution. In particular, when the initial state satisfies $r_1=0$, the Lyapunov stability theorem guarantees that the system state always stays in the equivalence class of $|\psi_f\rangle$, i.e., system (\ref{e1}) has a unique solution.

Next, we discuss the case when the initial state is in $\mathcal{O}$.
In this case, all quantum states satisfying $\cos\phi=0$ and $r_1\ne0$ form a longitude circle with $\phi=\frac{\pi}{2}$ and $\frac{3\pi}{2}$. For each quantum state $|\psi_q\rangle$ in $\mathcal{O}$, there exists $t_1$ such that the relative phase $\phi$ changes from $\phi(t_q)=\frac{\pi}{2}+p\pi$ to $\phi(t)\neq\frac{\pi}{2}+p\pi\;(p=0,1)$ in $(t_q,t_1]$.
In the Bloch sphere, the system trajectory intersects the longitude circle with $\phi=\frac{\pi}{2}$ and $\frac{3\pi}{2}$ in a non-overlap and non-tangent way in $[t_q,t_1]$, i.e., the vector field of the two-level system (\ref{e1}) is transversal to the non-Lipschitz set
$\mathcal{O}$. Hence, the system (\ref{e1}) in this case has a unique solution for every initial condition in $\mathcal{O}$ \cite{kawski1989stabilization}.
\quad $\blacksquare$
\end{pf}

\begin{rem}\label{7}
The concept of transversality is involved in the proof of Theorem \ref{6}, which is a description on how two objects intersect. For two intersecting curves, if they are not tangent, then they are said to be transversal each other. Readers can refer to \cite{golubitsky2012stable} for more general concepts and criteria for transversality.
\end{rem}

\section{Analysis of finite-time stability of two-level quantum control systems}\label{sec4}
For two-level quantum systems, we have the following finite-time stability theorem.
\begin{thm}\label{8}
Under the action of the controller (\ref{e35}), the system (\ref{e1}) with the Hamiltonians in (\ref{e22e23}) is globally finite-time stable, that is, the system will be stabilized in the equivalence class of the target state $|\psi_f\rangle=|1\rangle$ within a finite time.
\end{thm}
\begin{pf}
With (\ref{e28}), system (\ref{e1}) can be written as
\begin{equation}\label{e38}
{{\left[\!\begin{array}{c}\dot{r_1}e^{i\phi_a}+i r_{1}e^{i\phi_a}\dot{\phi}_a\\
\dot{r_2}e^{i\phi_b}+i r_{2}e^{i\phi_b}\dot{\phi}_b\end{array}\!\right]}\!=\!
-i{\left[\!\begin{array}{c}r_{1}e^{i\phi_a}\\-r_{2}e^{i\phi_b}\end{array}\!\right]}
\!-i u_1{\left[\!\begin{array}{c}-i r_{2}e^{i\phi_b}\\i r_{1}e^{i\phi_a}\end{array}\!\right]}},
\end{equation}
which is equivalent to the following relation
\begin{equation}\label{e40}
\left\{
   \begin{array}{l}
   \dot{r}_{1}=-u_{1}r_{2}\cos\phi=-Kr_{1}^{\alpha}r_{2}|\cos\phi|^{\alpha+1},\\
   r_{1}\dot{\phi}_a=-r_1-u_{1}r_{2}\sin\phi=-r_1-Kr_{1}^{\alpha}r_{2}|\cos\phi|^{\alpha}\sin\phi,\\
   \dot{r}_{2}=-u_{1}r_{1}\cos\phi=K|r_{1}\cos\phi|^{\alpha+1},\\
   r_{2}\dot{\phi}_b=r_2-u_{1}r_{1}\sin\phi=r_2-Kr_{1}^{\alpha+1}|\cos\phi|^{\alpha}\sin\phi.
   \end{array}
\right.
\end{equation}
Theorem \ref{6} implies that system (\ref{e40}) also has a unique solution. Therefore, $|\cos\phi|^{\alpha+1}$ in (\ref{e40}) can be regarded as a function of $t$, denoted as $g(t)$. Thus, we have
\begin{equation}\label{e41}
\left\{
   \begin{array}{l}
   \dot{r}_{1}=-Kr_{1}^{\alpha}r_{2}g\left(t\right),\\
   \dot{r}_{2}=Kr_{1}^{\alpha+1}g\left(t\right).
   \end{array}
\right.
\end{equation}
The objective is to stabilize the state $[r_1,\,r_2]^\mathrm{T}$ of system \eqref{e41} to the target point $[0,\,1]^\mathrm{T}$ from the initial point $[r_1(0),\,r_2(0)]^\mathrm{T}$.
Since $r_{1}^{2}+r_{2}^{2}=1$, we only need to consider whether the controlled variable $r_1$ defined on $\mathbb{R}_+$ can be stabilized to the origin $0$ from the initial point $r_1(0)$. Expressing $r_2$ with $r_1$, we have
\begin{equation}\label{e42}
r_2=\left(1-r_{1}^{2}\right)^{\frac{1}{2}}
=1-\sum_{j=1}^\infty\frac{\mathrm{C}_{2j}^{j}}{2^{2j}\times\left(2j-1\right)}r_{1}^{2j}.
\end{equation}
Substituting (\ref{e42}) into the first equation of (\ref{e41}) gives
\begin{equation}\label{e43}
\begin{aligned}
\dot{r}_1=&-Kr_{1}^{\alpha}g\left(t\right)+\sum_{j=1}^\infty\frac{\mathrm{C}_{2j}^{j}r_{1}^{2j}Kr_{1}^{\alpha}g\left(t\right)}{2^{2j}\times\left(2j-1\right)}.
\end{aligned}
\end{equation}
For convenience of analysis, we write (\ref{e43}) as
\begin{equation}\label{e44}
\dot{r}_1=f\left(r_1\right)=p_0(r_1)+\sum_{j=1}^\infty p_{j}\left(r_1\right)=\sum_{j=0}^\infty p_{j}\left(r_1\right),
\end{equation}
where $p_0(r_1)=-Kr_{1}^{\alpha}g(t)$ and $p_{j}(r_1)=\frac{\mathrm{C}_{2j}^{j}Kr_{1}^{\alpha+2j}g(t)}{2^{2j}\times(2j-1)}$ $(j\ge 1)$.

In what follows, we prove that system (\ref{e44}) is finite-time stable.
The proof can be divided into two steps.\\
\textbf{Step 1} The system defined by
\begin{equation}\label{e45}
{\dot{r}_1=p_0\left(r_1\right)}
\end{equation}
is finite-time stable.\\
\textbf{Step 2} The system (\ref{e44}) is globally finite-time stable.

\textbf{Proof of Step 1.}
According to Lemma \ref{A6} in the Appendix, to prove the finite-time stability of system (\ref{e45}), we only need to verify that system (\ref{e45}) is asymptotically stable and has a negative degree of homogeneity.

\emph{Asymptotic stability.}  For the Lyapunov function $V(r_1)=r_1^2$, we calculate its Lie derivative along the trajectory of system (\ref{e45}) and have
\begin{equation}\label{e46}
\begin{aligned}
L_{p_0}V\left(r_1\right)&=\langle\nabla V\left(r_1\right),p_0\left(r_1\right)\rangle\\
&=2r_{1}p_0\left(r_1\right)=-2Kr_{1}^{\alpha+1}g\left(t\right).
\end{aligned}
\end{equation}
From (\ref{e46}), the Lyapunov function $V(r_1)$ is non-increasing and $L_{p_0}V(r_1)$ is bounded. The fact that $L_{p_0}V(r_1)$ is bounded implies that $L_{p_0}V(r_1)$ is uniformly continuous, and therefore the Barbalat's lemma \cite{khalil2002nonlinear} guarantees that $L_{p_0}V(r_1)\rightarrow0$ as $t\rightarrow\infty$. Considering $g(t)>0$, we have $r_1\rightarrow0$, that is, system (\ref{e45}) is asymptotically stable.

\emph{Negative degree of homogeneity.}  According to Definition \ref{A3} in the Appendix, when $0<\alpha<1$ and the dilation is taken as $\delta_\varepsilon^1$, the vector field $p_0(r_1)$ satisfies
\begin{equation}\label{e47}
{p_0\left(\varepsilon r_1\right)=\varepsilon^{\alpha}p_0\left(r_1\right)=\varepsilon^{1+\left(\alpha-1\right)}p_0\left(r_1\right)}.
\end{equation}
Therefore, the degree of homogeneity of the vector field $p_0(r_1)$ with respect to the dilation $\delta_\varepsilon^1$ is $k_0=\alpha-1<0$. It follows from Lemma \ref{A6} in the Appendix that the origin of system (\ref{e45}) is finite-time stable.

\textbf{Proof of Step 2.}
For $j=1,2,3,\ldots$, we calculate the degree of homogeneity of the vector field
$p_j(r_1)$ in (\ref{e44}) with respect to the dilation $\delta_\varepsilon^1$ and $k_{j}$. We have
\begin{equation}\label{e48}
\begin{aligned}
p_j\left(\varepsilon
r_1\right)&=\frac{\mathrm{C}_{2j}^{j}}{2^{2j}\times\left(2j-1\right)}K\varepsilon^{\alpha+2j}r_{1}^{\alpha+2j}g\left(t\right)\\
&=\varepsilon^{1+\left(\alpha+2j-1\right)}p_j\left(r_1\right)\\
&=\varepsilon^{1+k_{j}}p_j\left(r_1\right).
\end{aligned}
\end{equation}
That is, $k_{j}=\alpha+2j-1$ $(j=1,2,3,\ldots)$.

Note that the degree of homogeneity of $V(r_1)$ with respect to the dilation $\delta_\varepsilon^1$ is $l_1=2$, $\langle\nabla V(r_1),p_{j}(r_1)\rangle$ $(j=0,1,2,\ldots)$ is continuous and its degree of homogeneity with respect to $\delta_\varepsilon^1$ is $l_1+k_{j}$. We take $V_1=V(r_1)$ and $V_2=\langle\nabla V(r_1),p_j(r_1)\rangle$ for Lemma \ref{A5} in the Appendix. Since $l_1=2>0$ and $l_2=l_1+k_{j}=\alpha+2j+1>0$, Lemma \ref{A5} implies
\begin{equation}\label{e49}
\langle\nabla V\left(r_1\right),p_j\left(r_1\right)\rangle\leq-c_{j}V\left(r_1\right)^{\frac{\alpha+2j+1}{2}},
\end{equation}
where $c_{j}\!=\!-\max_{\{r_1:V(r_1)=1\}}\langle\nabla V(r_1),p_j(r_1)\rangle\!\in\mathbb{R}$ $(j=0,1,2,\ldots)$.
Thus,
\begin{equation}\label{e50}
\begin{aligned}
&\langle\nabla V\left(r_1\right),f\left(r_1\right)\rangle\\
\leq&-c_{0}V\left(r_1\right)^{\frac{\alpha+1}{2}}-\cdots-c_{j}V\left(r_1\right)^{\frac{\alpha+2j+1}{2}}-\cdots\\
=&\;V\left(r_1\right)^{\frac{\alpha+1}{2}}\big(-c_0+\mathcal{U}(r_1)\big),
\end{aligned}
\end{equation}
where $\mathcal{U}(r_1)\triangleq-c_{1}V(r_1)^{\frac{2}{2}}-\cdots-c_{j}V(r_1)^{\frac{2j}{2}}-\cdots$. Since $\frac{2j}{2}>0$ for $j\geq1$, $\mathcal{U}(r_1)$
is a continuous function with $\mathcal{U}(0)=0$.

Now, we show that \eqref{e50} satisfies the condition in (\ref{e16}).
Assume that there exists an open neighborhood $\mathcal{V}$ of the origin such that $\mathcal{U}(r_1)<\frac{c_0}{2}$ holds for any $r_1\in\mathcal{V}$. Then, (\ref{e50}) can be written as
\begin{equation}\label{e51}
\langle\nabla V\left(r_1\right),f\left(r_1\right)\rangle<-\frac{c_0}{2}V\left(r_1\right)^{\frac{\alpha+1}{2}},
\end{equation}
where $c_0>0$ and $\frac{\alpha+1}{2}\in(0,1)$. Thus, the condition (\ref{e16}) in Theorem \ref{5} is satisfied. In view of the positive definiteness of $V(r_1)$, Theorem \ref{5} guarantees that the origin is a finite-time stable equilibrium point of system (\ref{e44}).

Next, we verify the existence of the open neighborhood $\mathcal{V}$, that is, there exist $r_1$ such that $\mathcal{U}(r_1)<\frac{c_0}{2}$ holds. Considering that $c_{j}=-\max_{\{r_1:V(r_1)=1\}}\langle\nabla V(r_1),p_j(r_1)\rangle$ $(j=0,1,2,\ldots)$ and $r_1=1$ holds when $V(r_1)=1$, we calculate $c_0$ and $c_{j}$ $(j\ge 1)$ as
\begin{equation}\label{e52}
{c_0=-\langle\nabla V\left(r_1\right),p_0\left(r_1\right)\rangle=2Kg\left(t\right)},
\end{equation}
\begin{equation}\label{e53}
c_{j}=-\langle\nabla V\left(r_1\right),p_j\left(r_1\right)\rangle
=-\frac{2K\mathrm{C}_{2j}^{j}g(t)}{2^{2j}\times\left(2j-1\right)}.
\end{equation}

It follows from \eqref{e53} that
\begin{equation}\label{e54}
\begin{aligned}
&\mathcal{U}\left(r_1\right)=-c_1V\left(r_1\right)^{\frac{2}{2}}-\cdots-c_{j}V\left(r_1\right)^{\frac{2j}{2}}-\cdots\\
=&\,2Kg(t)\Big[\frac{1}{2}V(r_1)+\cdots+\frac{\mathrm{C}_{2j}^{j}V(r_1)^j}{2^{2j}\times\left(2j-1\right)}+\cdots\Big]\\
=&\,2Kg(t)\big[1-\left(1-V(r_1)\right)^{\frac{1}{2}}\big].
\end{aligned}
\end{equation}
Substituting (\ref{e54}) and \eqref{e52} into $\mathcal{U}(r_1)<\frac{c_0}{2}$, we have $r_1<\frac{\sqrt{3}}{2}$, that is, $\mathcal{V}=\{r_1:r_1<\frac{\sqrt{3}}{2}\}$. This shows the existence of $\mathcal{V}$ in (\ref{e51}). Furthermore, since all moments $t_q$ corresponding to $|\psi_q\rangle$ constitute a zero measure set and any other state different from $|\psi_q\rangle$ satisfies $\dot{V}<0$, $r_1$ can always converge into $\mathcal{V}$ within a finite time for every initial state $r_1(0)\notin\mathcal{V}$.

We can conclude that the origin of system (\ref{e44}) is a global finite-time stable equilibrium point. That is, $r_1$ can be stabilized to the origin within a finite time. Equivalently, the quantum state is stabilized to the equivalence class of the target state $|\psi_f\rangle=|1\rangle$ within a finite time. \quad $\blacksquare$
\end{pf}

\begin{rem}\label{9}
According to the proof of Theorem \ref{8}, (\ref{e51}) always holds for system (\ref{e44}). From Theorem \ref{5}, the settling time function satisfies $T(r_1(0))<\frac{4}{c_{1}(1-\alpha)}V(r_1(0))^{\frac{1-\alpha}{2}}$ when $r_1(0)\in\mathcal{V}$. When $r_1(0)\notin\mathcal{V}$, the calculation of the settling time relies on the system equation (\ref{e44}), and therefore it is not easy to give a analytical bound for $T(r_1(0))$. In addition, since $c_1$, $V(r_1(0))$, and $\alpha\in(0,1)$ are bounded, $T(r_1(0))$ is also bounded although the bound may vary with $\alpha$.
\end{rem}

\section{Numerical examples}\label{sec5}
\begin{figure}
\begin{center}
\includegraphics[width=9cm]{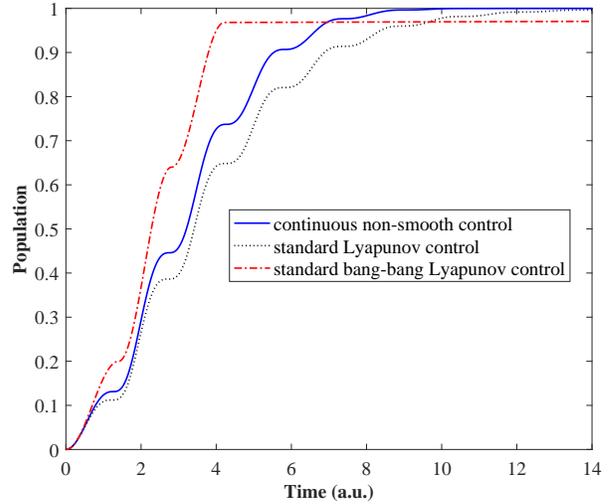}    
\caption{The population evolution of the target state for $|\psi(0)\rangle=|0\rangle$ under the continuous non-smooth control $u_1$, the standard Lyapunov control $u_1^s$, and the standard bang-bang Lyapunov control $u_1^b$.}  
\label{fig2}                                 
\end{center}                                 
\end{figure}

\begin{figure}
\begin{center}
\includegraphics[width=9cm]{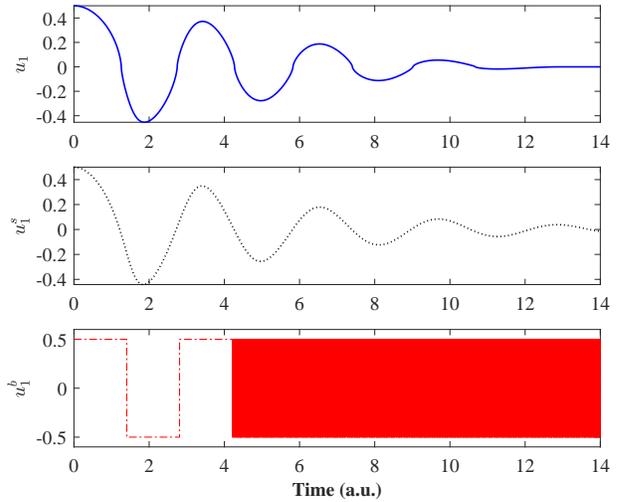}    
\caption{The continuous non-smooth control $u_1$, the standard Lyapunov control $u_1^s$, and the standard bang-bang Lyapunov control $u_1^b$ for $|\psi(0)\rangle=|0\rangle$.}  
\label{fig3}                                 
\end{center}                                 
\end{figure}

We choose a spin-$\frac{1}{2}$ system to present numerical results. Spin-1/2 systems have wide applications in e.g., quantum computation, quantum sensing and quantum control \cite{wang2020identifiability}.
In simulation, we set $K=0.5$ and $\alpha=\frac{2}{3}$ for the control law in (\ref{e35}). We also consider simulation results under the standard Lyapunov control law $u_1^s$ \cite{shuang2007quantum} and the standard bang-bang Lyapunov control law $u_1^b$ \cite{kuang2017rapid} for comparison, where $u_1^s$ and $u_1^b$ are
\begin{gather}
{u_1^s=K\Im\big[e^{i\angle{\langle\psi|\psi_f\rangle}}{\langle\psi_f|H_1|\psi\rangle}\big]},\label{e55}\\
{u_1^b=K\textrm{sign}\big(\Im\big[e^{i\angle{\langle\psi|\psi_f\rangle}}{\langle\psi_f|H_1|\psi\rangle}\big]\big)}.\label{e56}
\end{gather}
We choose the initial state as $|\psi(0)\rangle=|0\rangle\notin\mathcal{V}$ and $K=0.5$ for the standard Lyapunov control law $u_1^s$ and the standard bang-bang Lyapunov control law $u_1^b$. The simulation results are shown in Fig. \ref{fig2} and Fig. \ref{fig3}.

According to Fig. \ref{fig2} and simulation data, the settling time associated with the initial state $|0\rangle$ can be obtained as $t_f\approx11.6270$ a.u.. At $t=11.6270$ a.u., the populations of the target state under $u_1^s$, $u_1^b$, and $u_1$ are 0.9902, 0.9699, and 1.0000, respectively. It can be seen from Fig.~\ref{fig3} that the control law in this paper is indeed continuous and non-smooth while the standard Lyapunov control is smooth and the standard bang-bang Lyapunov control is discontinuous with chattering.

We further perform simulation experiments for the initial state $|\psi(0)\rangle=[\frac{1}{2},\,\frac{\sqrt{3}}{2}]^\mathrm{T}\in\mathcal{V}$. The simulation data indicate that the settling time is  $t_f\approx7.5$ a.u.. According to Remark \ref{9}, the settling time associated with the initial state $|\psi(0)\rangle=[\frac{1}{2},\,\frac{\sqrt{3}}{2}]^\mathrm{T}$ satisfies $T(\frac{1}{2})\approx7.5\,\mathrm{a.u.}<\frac{4}{c_1(1-\alpha)}V(\frac{1}{2})^{\frac{1-\alpha}{2}}=9.52 \,\mathrm{a.u.}$, which is consistent with the theoretical result.

\begin{rem}\label{10}
For the numerical example in this section, it follows from \cite{boscain2006time} that the minimum transfer time $T_O$ from $|0\rangle$ to $|1\rangle$ satisfies $1.9505<T_O<6.1655$. In the minimum time control scheme, the control law is an optimal bang-bang control and takes the maximal admissible value either 0.5 or -0.5 at each moment during the whole control process. While in the finite-time Lyapunov control scheme, the control law is a continuous non-smooth control and only takes the maximal admissible amplitude 0.5 at the initial moment. The settling time is longer than the minimum time $T_{O}$.
\end{rem}

\section{Conclusion}\label{sec6}
We investigated the finite-time stability and presented a Lyapunov stability criterion for finite-dimensional quantum systems. A new continuous non-smooth control law was proposed and the finite-time stabilization towards an eigenstate of the internal Hamiltonian was achieved for two-level quantum systems. Based on the transversality condition in the Bloch space, we proved the uniqueness of solutions of the system with the continuous non-smooth controller. Using the finite-time Lyapunov stability theory and the homogeneity theorem, we also proved the finite-time stability of the control system. The effectiveness of the proposed continuous non-smooth control law was illustrated by numerical examples.
Future research includes optimizing the parameter $\alpha$ in the controller to achieve an optimal performance, and extending the finite-time control scheme to high-dimensional quantum systems and stochastic open quantum systems with measurement feedback.


\section*{Appendix: Homogeneity theory for finite-time stability}\label{Appendix}    
Several concepts and results related to homogeneity are listed here, which can be found in \cite{bhat2005geometric}.
\begin{defn} \label{A1}
Let $d=\left(d_1,d_2,\ldots,d_{n-1}\right)$ be a set of positive real numbers. For a set of coordinates  $r=\left(r_1,r_2,\ldots,r_{n-1}\right)$ in $\mathbb{R}^{n-1}$, define the dilation $\delta_\varepsilon^d$ of $r$ as the following coordinate vector
\begin{equation}\label{e57}
\begin{aligned}
&\delta_\varepsilon^d\left(r\right)=\left(\varepsilon^{d_1}r_1,\ldots,\varepsilon^{d_{n-1}}r_{n-1}\right),\forall\varepsilon>0
\end{aligned}
\end{equation}
where $d_j$ is the weight of the coordinate $r_j$. The dilation with $d_1=\cdots= d_{n-1}=1$ is called a standard dilation.
\end{defn}
\begin{defn} \label{A2}
A function $V:\mathbb{R}^{n-1}\rightarrow\mathbb{R}$ is said to be homogeneous of degree $m\left(m\in\mathbb{R}\right)$ with respect to $\delta_{\varepsilon}^{d}$ if
\begin{equation}\label{e58}
{V\left(\delta_\varepsilon^d(r)\right)=\varepsilon^{m}V(r),\;\forall r\in\mathbb{R}^{n-1},\forall\varepsilon>0.}
\end{equation}
\end{defn}
\begin{defn} \label{A3}
A vector field $f(r):\,\mathbb{R}^{n-1}\rightarrow \mathbb{R}^{n-1}$ with $f(r)=(f_1(r),\,\ldots,\,f_{n-1}(r))^\mathrm{T}$ is said to be homogeneous of degree $k\left(k\in\mathbb{R}\right)$ with respect to $\delta_{\varepsilon}^{d}$ if for each $i=1,\ldots,n-1$, $f_j$ is homogeneous of degree $k+d_j$, that is,
\begin{equation}\label{e59}
f_j\left(\delta_\varepsilon^d\left(r\right)\right)=\varepsilon^{k+d_{j}}f_{j}\left(r\right),\;\forall r\in\mathbb{R}^{n-1},\,\forall\varepsilon>0.
\end{equation}
\end{defn}
\begin{lem} \label{A4}
Assume that the function $f:\,\mathbb{R}^{n-1}\rightarrow \mathbb{R}^{n-1}$ is homogeneous of degree $k\left(k\in\mathbb{R}\right)$ with respect to $\delta_{\varepsilon}^{d}$ and the origin is a locally asymptotically stable equilibrium point. Then, when $m>\max\{-k,0\}$, there exists a Lyapunov function $V$ such that $V$ and its time derivative $\dot{V}$ are homogeneous of degrees $m$ and $m+k$ with respect to $\delta_{\varepsilon}^{d}$, respectively.
\end{lem}
\begin{lem} \label{A5}
Let $V_1$ and $V_2$ be continuous real-valued functions defined on $\mathbb{R}^{n-1}$ and $V_1$ be positive definite. Suppose that $V_1$ and $V_2$ are homogeneous of degrees $l_1>0$ and $l_2>0$ with respect to $\delta_{\varepsilon}^{d}$, respectively. Then, for every $r\in\mathbb{R}^{n-1}$, the following holds:
\begin{equation}\label{e60}
\begin{aligned}
&\Big(\min_{\{z:V_1(z)=1\}}V_2(z)\Big)\big(V_1(r)\big)^{\frac{l_2}{l_1}}
\leq V_2(r)\\
\leq&\;\Big(\max_{\{z:V_1(z)=1\}}V_2(z)\Big)\big(V_1(r)\big)^{\frac{l_2}{l_1}}.
\end{aligned}
\end{equation}
\end{lem}
The following lemma shows the application of the homogeneity theory to finite-time stability.
\begin{lem} \label{A6}
Let $f\left(r\right)=\left(f_1\left(r\right),\ldots,f_{n-1}\left(r\right)\right)^\mathrm{T}:\,\mathbb{R}^{n-1}$ $\rightarrow \mathbb{R}^{n-1}$ be a continuous vector function and be homogeneous of degree $k\,\left(k\in\mathbb{R}\right)$ with respect to $\delta_{\varepsilon}^{d}$, where $d=\left(d_1,d_2,\ldots,d_{n-1}\right)$ is a set of positive real numbers and $\varepsilon>0$. Then, the origin is a finite-time stable equilibrium point if and only if it is an asymptotically stable equilibrium point and $k<0$.
\end{lem}


\end{document}